\documentclass[a4paper,fleqn,usenatbib,useAMS,full]{mnras}

\pdfoutput=1

\newif\ifpdf
\ifx\pdfoutput\undefined
\pdffalse
\else
\pdfoutput=1
\pdftrue
\fi
\ifpdf
\usepackage{graphicx}
\usepackage{epstopdf}
\else
\usepackage{graphicx}
\fi
\usepackage{amsmath}	
\usepackage{amssymb}	
\usepackage{multicol}        
\usepackage{bm}		
\usepackage{pdflscape}	
\usepackage{caption}
\usepackage{subcaption}
\usepackage[T1]{fontenc}
\usepackage{ae,aecompl}
\usepackage{txfonts}

\title[Bispectrum Multipoles]{Information Content of the Angular Multipoles of Redshift-Space Galaxy Bispectrum}

\author[Praful Gagrani, Lado Samushia]{Praful Gagrani$^{1}$\thanks{Contact e-mail: \href{mailto:praful@phys.ksu.edu}{praful@phys.ksu.edu}},  Lado Samushia$^{1,2,3}$ \thanks{Contact e-mail: \href{mailto:lado@phys.ksu.edu}{lado@phys.ksu.edu}}
\\
$^{1}$Department of Physics, Kansas State University, 116 Cardwell Hall, Manhattan, KS, 66506, USA \\
$^{2}$National Abastumani Astrophysical Observatory, Ilia State University, 2A Kazbegi Ave., GE-1060 Tbilisi, Georgia \\
$^3$Institute of Cosmology \& Gravitation, University of Portsmouth, PO1 3FX, UK }

\date{Last updated 2016 June 22; in original form 2016 June 5}

\pubyear{2016}

\begin{document}
\label{firstpage}
\pagerange{\pageref{firstpage}--\pageref{lastpage}}
\maketitle

\begin{abstract}

The redshift-space bispectrum (three point statistics) of galaxies depends on
the expansion rate, the growth rate, and geometry of the Universe, and hence can
be used to measure key cosmological parameters. In a homogeneous Universe the
bispectrum is a function of five variables and unlike its two point statistics
counterpart -- the power spectrum, which is a function of only two variables --
is difficult to analyse unless the information is somehow reduced. The most
commonly considered reduction schemes rely on computing angular integrals over
possible orientations of the bispectrum triangle, thus reducing it to sets of
function of only three variables describing the triangle shape. We use Fisher
information formalism to study the information loss associated with this angular
integration. Without any reduction, the bispectrum alone can deliver constraints on the growth rate parameter $f$ that are better by a factor of $2.5$ compared to the power
spectrum, for a sample of luminous red galaxies expected from near future
galaxy surveys at a redshift of $z\sim0.65$. At lower redshifts the improvement could be
up to a factor of $3$. We find that most of the
information is in the azimuthal averages of the first three even multipoles. This suggests that the bispectrum of every
configuration can be reduced to just three numbers (instead of a 2D function)
without significant loss of cosmologically relevant information.\\
\end{abstract}


\section{Introduction}
\label{sec:intro}

The statistical properties of matter distribution in the Universe depend on its
expansion and growth history and can be used to measure key cosmological
parameters describing the composition of the Universe, the nature of dark
energy, and gravity.

The power spectrum (or its Fourier conjugate the correlation function) is
currently the most widely used statistical measurement for the purposes of
cosmological analysis of galaxy surveys. The power spectrum of matter is defined
as a two point statistics of a Fourier transformed overdensity field
$\delta(\bm{r})$,
\begin{equation}
    P(\bm{k}) \equiv
\frac{\left<|\delta(\bm{k})|^2\right>}{V_\mathrm{s}},
\end{equation}
where
\begin{equation}
\delta(\bm{k}) =
\displaystyle\int\!\mathrm{d}\bm{r}\,\delta(\bm{r})\mathrm{e}^{-\mathrm{i}\bm{k}\bm{r}},
\end{equation}
brackets denote ensemble average, and $V_\mathrm{s} \equiv
\displaystyle\int\!\mathrm{d}\bm{r}$ is the observed volume.  

For a statistically isotropic field the power spectrum would only depend on the
magnitude of the wavevector, $k=|\bm{k}|$.  The observed galaxy field is
however anisotropic with respect to the line-of-sight direction to the observer,
mainly due to the redshift-space distortions \citep[RSD,][]{kaiser1987clustering} and the Alcock-Paczinsky
effects \citep[AP,][]{alcock1979evolution}. Because of this anisotropy, in addition to the magnitude of the
wavevector $k$, the power spectrum also depends on its angle with respect to the
line-of-sight $\theta$, making it a function of two variables.
 
To make the cosmological analysis numerically less demanding the power spectrum
is usually reduced to the coefficients of the Legendre-Fourier expansion with
respect to $\mu$=$\cos(\theta)$ \citep{taylor1996non}
\begin{equation}
\label{eq:Pell}
P_\ell(k) \equiv \frac{2\ell+1}{2}\displaystyle\int\limits_{-1}^1\!\mathrm{d}\mu\,P(k,\mu)\mathcal{L}_\ell(\mu).
\end{equation}
Recent studies showed that the first three even Legendre coefficients contain
almost all of the information on key cosmological parameters. This suggest that
for the purposes of cosmological analysis the power spectrum at each wavevector
can be replaced just by three numbers (instead of a function of $\mu$) without
a significant loss of information \citep{taruya2011forecasting, kazin2010baryonic, beutler2014clustering}.

The bispectrum (or its Fourier conjugate the three-point correlation function),
defined as,
\begin{equation}
B(\bm{k}_1,\bm{k}_2,\bm{k}_3) \equiv \frac{\langle\delta(\bm{k}_1)\delta(\bm{k}_2)\delta(\bm{k}_3)\rangle}{V_\mathrm{s}}
\end{equation} 
is more difficult to measure and to model, and is not currently used as
frequently as the power spectrum to derive cosmological constraints \citep{song2015cosmology, greig2013cosmology, scoccimarro1999bispectrum, sefusattikomatsu}. 
The bispectrum measurements have mostly been considered as a means of estimating the primordial
non-Gaussianity in the matter field \citep{tellarini, sefusatti2012halo}, but a number of recent studies used them for BAO and RSD constraints \citep{slepian2016modeling, slepian2016detection, gil2015power, gil2016clustering}. 

If the statistical properties of the Universe are homogeneous (a key assumption in the
standard model of cosmology) the bispectrum is non-zero only for
$\bm{k}_1+\bm{k}_2+\bm{k}_3=0$ ($\bm{k}$ vectors must make a triangle) reducing
the number of variables from nine to six. From now on we will write
$B(\bm{k}_1,\bm{k}_2)$ assuming the third vector to be equal to $\bm{k}_3 =
-\bm{k}_1 - \bm{k}_2$. The partial isotropy with respect to rotations around the
line-of-sight axis removes one more variable, making the bispectrum a five
dimensional function.  One possible choice of these five variables is a triplet
$k_1, k_2, k_3$ ($k_i\equiv|\bm{k}_i|$), describing the shape of the bispectrum
triangle and two angles describing its orientation, e.g.  $\theta_1$ -- the
angle of $\bm{k}_1$ vector with respect to the line-of-sight direction, and
$\xi$ -- azimuthal angle of $\textbf{k}_2$ around $\textbf{k}_1$ (see
Sec.~\ref{sec:review} for a formal definition).

An obvious extension of the Legendre-Fourier decomposition of the power spectrum
is a spherical harmonics decomposition of the bispectrum for angles $\theta_1$
and $\xi$ \citep{scoccimarro2015fast}. Unlike the power spectrum, this double angular multipole expansion of the bispectrum does not truncate at finite order (see Sec.~\ref{sec:multipoles}). The main objective of this work is to identify the expansion coefficients that contain the most cosmologically relevant information (see Sec.~\ref{sec:fisher}).

Galaxies provide a biased, discrete sampling of the underlying matter field and
along with the cosmic microwave background experiments currently provide one of
the best estimates of the clustering of matter in the Universe \citep{ade2014planck, schlegel2009bigboss}. Our Fisher
information based computations suggest the five dimensional bispectrum with no
reduction can deliver up to factor of $1.2$ better constraints on the growth
rate parameter $f$ compared to the power spectrum, from a sample of emission
line galaxies (ELG) expected from future surveys such as the Dark Energy
Spectroscopic Instrument survey (DESI; \cite{levi2013desi}) and Euclid satellite surveys \citep{laureijs2011euclid}  at a
redshift of $z\sim1$ (see Sec.~\ref{sec:results}). For a sample of Luminous Red Galaxies (LRG) at lower redshifts the improvement could
be as large as a factor of 3.

We show that most of this information is contained in the first three even
multipoles in angle $\theta_1$ averaged over $\xi$. Constraints on key 
cosmological parameters from these multipoles are weaker compared to the 
constraints derived from the full bispectrum by no more than 10 per cent at all 
redshifts and for all tracer types we studied. This suggests that
a bispectrum of each triangular configuration can be replaced by just three
numbers (as opposed to a two variable function) for all practical purposes (see
Sec.~\ref{sec:conclusions}). 

%
%

\section{Review of Power Spectrum and Bispectrum}
\label{sec:review}
\subsection{Leading Order Model}
We will start with a standard assumption that galaxies form a Poisson sample of
a biased matter density field \citep{peebles1980large},
\begin{equation}
n(\textbf{x}) =
\bar{n}\left[1+b_1\delta(\textbf{x})+\frac{b_2}{2}\delta(\textbf{x})^2\right],
\end{equation}
where $b_1$ and $b_2$ are the first and second order bias parameters and we
ignore higher order bias terms as well as non-local contributions of
$\delta(\bm{x})$ to the number density of galaxies.

To the leading order in $\delta$ the power spectrum is given by
\citep{kaiser1987clustering},
\begin{equation}
\label{eq:Plin}
P(\textbf{k}) = (b_1+f \mu^2)^2P_\mathrm{m}(k),
\end{equation}
where $f$ is a growth rate and $P_\mathrm{m}$ is a one dimensional matter power
spectrum function that can be numerically computed for any cosmological 
model.\footnote{The bias and the growth rate can not be decoupled from the 
amplitude parameter $\sigma_8$ when using only the galaxy clustering data on 
linear scales at a single redshift. For brevity, we will continue using $b$ and 
$f$ to denote parameter combinations $b\sigma_8$ and $f\sigma_8$.}
Also in the leading order of perturbation theory and assuming local bias the
bispectrum of galaxies is given by \citep{scoccimarro2000bispectrum},
\begin{align}
\label{eq:bisp}
B(\textbf{k}_{1},\textbf{k}_{2},\textbf{k}_{3}) = &2\textbf{Z}_1(\mu_1)\textbf{Z}_1(\mu_2)\textbf{Z}_2P(k_1)P(k_2)\\
\nonumber
&+ \text{cyclic terms},
\end{align}
\noindent
where
\begin{align}
&\textbf{Z}_1(\mu) = \left(b_1+f \mu^2\right),\\
\nonumber
&\textbf{Z}_2 = \left\{\frac{b_2}{2} + b_1F_2(\textbf{k}_{1},\textbf{k}_{2}) +
f\mu_3^2G_2(\textbf{k}_{1},\textbf{k}_{2})\right. \\
&-
\left.\frac{f\mu_3k_3}{2}\left[\frac{\mu_1}{k_1}(b_1+f\mu^2_2)+\frac{\mu_2}{k_2}(b_1+f\mu^2_1)\right]\right\},
\end{align}
\begin{align}
&F_2(\textbf{k}_{1},\textbf{k}_{2}) = \frac{5}{7} + \frac{\textbf{k}_{1}. \textbf{k}_{2}}{2k_1k_2}\bigg(\frac{k_1}{k_2}+\frac{k_2}{k_1}\bigg)+\frac{2}{7}\bigg(\frac{\textbf{k}_{1}. \textbf{k}_{2}}{k_1k_2}\bigg)^2, \\
&G_2(\textbf{k}_{1},\textbf{k}_{2}) = \frac{3}{7} + \frac{\textbf{k}_{1}. \textbf{k}_{2}}{2k_1k_2}\bigg(\frac{k_1}{k_2}+\frac{k_2}{k_1}\bigg)+\frac{4}{7}\bigg(\frac{\textbf{k}_{1}. \textbf{k}_{2}}{k_1k_2}\bigg)^2, 
\end{align}
\noindent
and  cyclic terms can be derived by
replacing indexes 1 and 2 in the first term by 2 and 3, and 1 and 3
respectively.

The AP effect induces distortions in
the measured power spectrum and the bispectrum that can be modeled by
substituting
\begin{align}
        &k \rightarrow \frac{k}{\alpha_\perp}\sqrt{1+\mu^2(A^{-2}-1)}\\
        &\mu \rightarrow \frac{\mu}{\sqrt{A^2 + \mu^2(1-A^2)}}
\end{align} 
and renormalizing the power spectrum by a factor of
$1/\alpha_\bot^2\alpha_\parallel$ and the bispectrum by the square of the same
factor.  $A = \alpha_\parallel / \alpha_\perp$ in the above equations and the
$\alpha$ parameters can be linked to properties of dark energy
\citep{ballinger1996measuring, simpson2010difficulties, samushia2011effects}. 

A standard practice when analysing galaxy power spectrum is to assume that the shape of
the matter power spectrum is well determined from external cosmological data
sets (e.g. the cosmic microwave background experiments) and to treat it as a function of four cosmological parameters $b_1, f,
\alpha_\perp, \alpha_\parallel$ The bispectrum in addition will depend on the
second order bias parameter $b_2$. For simplicity we ignore the commonly included 
$\sigma_\mathrm{FOG}$ \citep{jackson1972critique} parameter here. Its effect is to reduce information content on small scales. Since we are interested only on the 
relative constraining power of the power spectrum, the bispectrum, and their 
multipoles, this omission does not effect our main results. \footnote{When fitting real data more ``nuisance'' parameters are required to effectively describe the shortcomings of theoretical modelling. We ignore the effect of these ``nuisance'' parameters here as well since they depend on the specifics of modelling and do not effect our main results anyway.} These parameters then can be estimated from
the measured power spectrum and the bispectrum. We will adhere to this standard
assumption and will ignore other cosmological parameters that may be relevant
(e.g. $f_\mathrm{NL}$ describing primordial non-Gaussianity, or $N_\mathrm{eff}$
number of neutrino species).
    
\subsection{Variance of the Measurements}
    
If a power spectrum is measured from an observed volume $V_\mathrm{s}$ using
optimal estimators \citep{feldman1993power} the variance of the measurement is  
\begin{equation}\label{eq:Pcov}
\langle\left[ \Delta P(\textbf{k})\right]^2\rangle = \bigg(P(\textbf{k}) +
\frac{1}{\bar{n}}\bigg)^2,
\end{equation}
where $\Delta P$ is the difference between the true power spectrum and the one
estimated from finite (and noisy) data and $\overline{n}$ is the average number 
density of galaxies. In an analogous way, for the bispectrum
measured with an optimal estimator the variance is
\citep{scoccimarro2000bispectrum, sefusatti2006cosmology}
\begin{equation}
\label{eq:Bcov}
 \langle \left[\Delta B(\textbf{k}_1,\textbf{k}_2)\right]^2\rangle =
V_\mathrm{s}\left(P(\textbf{k}_1) +
\frac{1}{\overline{n}}\right)
\left(P(\textbf{k}_2)+\frac{1}{\overline{n}}\right)
 \left(P(\textbf{k}_3) + \frac{1}{\overline{n}}\right).
\end{equation}

%
%

\section{Bispectrum Multipoles}
\label{sec:multipoles}
\subsection{Parametrization of the Bispectrum}
Eq.~(\ref{eq:Plin}) shows that the power spectrum can be expressed as a function
of only two variables -- $k$ and $\mu$. This results from the azimuthal symmetry
of the field and is true even when the linear theory expression in
Eq.~\eqref{eq:Plin} is replaced by its non linear equivalent.

Similarly, even though the bispectrum in Eq.~(\ref{eq:bisp}) is written in terms
of three vectors $\textbf{k}_{1}$, $\textbf{k}_{2}$ and $\textbf{k}_{3}$, as
discussed in Sec.~\ref{sec:intro}, because of various symmetries, only five
variables are in fact independent. Following \cite{scoccimarro2015fast} we
choose these variables to be the lengths of three wavevectors $k_1$, $k_2$,
$k_3$ -- describing the shape of the bispectrum triangle, and two angles
describing its orientation -- the angle $\theta_1$ of wavevector $\bm{k}_1$ with
respect to the line-of-sight direction, and the azimuthal angle $\xi$ of vector
$\bm{k}_2$ around $\bm{k}_1$. The first four variables are trivially obtained
from the original wavevectors while the $\xi$ can be computed from
\begin{equation} 
\mu_2 = \cos (\theta_1 ) \cos (\phi_{12} ) - \sin (\theta_1 ) \sin
(\phi_{12} )\cos (\xi ),
    \end{equation} 
where $\phi_{12}$ is the angle between $\textbf{k}_1$ and $\textbf{k}_2$,
\begin{equation}
\phi_{12} = \cos ^{-1}\left(\frac{\textbf{k}_1\textbf{k}_2}{k_1 k_2 }\right)  .
\end{equation}

\subsection{Series Expansion of Bispectrum}

The power spectrum can be decomposed into Legendre-Fourier series in angle $\mu$
\begin{equation}
P(\bm{k}) = \displaystyle\sum\limits_\ell P_\ell(k)\mathcal{L}_\ell(\mu)
\end{equation}
where $\mathcal{L}_\ell$ are Legendre polynomials of order $\ell$ and the
coefficients of decomposition can be found using Eq.~(\ref{eq:Pell}). In linear
theory only the first three even coefficients are nonzero and they contain most
of the information on key cosmological parameters. 

Since $0<\theta_1<\pi$ and $0\leq \xi<2\pi$, the bispectrum can be decomposed in
spherical harmonics of $\theta_1$ and $\xi$
\begin{equation}\label{eq:BispAng}
 B(k_1,k_2,k_3,\theta_1,\xi) =
\displaystyle\sum_{\ell}\!\displaystyle\sum_{m=-\ell}^{\ell}\! B_{\ell
m}(k_1,k_2,k_3)Y^m_\ell(\theta_1,\xi).
\end{equation}
Subsequently,
\begin{equation}\label{eq:BispEst}
B_{\ell m}(k_1,k_2,k_3) = \displaystyle\int\limits_{-1}^1
\!\mathrm{d}\cos(\theta)\!\displaystyle\int\limits^{2\pi}_0 \!\!d\xi
B(k_1,k_2,k_3,\theta_1,\xi) \, Y^{m*}_\ell(\theta_1,\xi).
\end{equation}

Unlike the power spectrum, the bispectrum multipole expansion does not terminate
at final $\ell$. Neither does it have zero odd multipoles.  Reducing bispectrum
to a finite number of its angular multipoles significantly simplifies the cosmological analysis.
This reduction however will inevitably result in a loss of information.

From the practical point of view, computing multipoles with $m=0$ is especially
simple \citep{scoccimarro2015fast}. It is therefore interesting by how much the information degrades
further if we only use $m=0$ multipoles in the analysis. We will show that the
loss of information associated with ignoring $m$ larger than zero is negligible.

We will also show that almost all of the information on key cosmological parameters
(compared to using the full bispectrum) is contained within the first three even
multipoles ($\ell = 0,2,4$ with $m=0$) of the bispectrum.

\subsection{Covariance of Bispectrum Multipoles}

The bispectrum multipoles from real data can be computed by summing over all
triangles with fixed values of $k_i$ and angular weights of Eq.~ (\ref{eq:BispEst}). This
corresponds to
\begin{align}
\nonumber
&\overline{B}_{\ell m}(k_1',k_2',k_3') \equiv\\ \nonumber
&\hspace{20pt}\frac{1}{2\pi}\displaystyle\int\mathrm{d}\bm{k}_1\mathrm{d}\bm{k}_2\frac{\delta(\bm{k}_1)\delta(\bm{k}_2)\delta(\bm{k}_3)}{V_\mathrm{s}}
Y^{m*}_\ell(\theta_1,\xi)\\ \nonumber
&\hspace{20pt}\times\hspace{-18pt}\quad\quad\frac{\delta^\mathrm{D}(k_1-k_1')}{k_1}\frac{\delta^\mathrm{D}(k_2-k_2')}{k_2}\frac{\delta^\mathrm{D}(k_3-k_3')}{k_3}=\\
&\hspace{20pt}\frac{1}{2\pi V_\mathrm{s}}
\displaystyle\int\mathrm{d}\theta_1\mathrm{d}\xi\mathrm{d}\phi_1
\delta(\bm{k}_1')\delta(\bm{k}_2')\delta(\bm{k}_3') Y^{m*}_\ell(\theta_1,\xi),
\end{align}
where we used the transformation of coordinates
\begin{align}
\nonumber
\mathrm{d}\bm{k}_1\mathrm{d}\bm{k}_2 &=
k_1^2\mathrm{d}k_1\mathrm{d}\cos(\theta_1)\mathrm{d}\phi_1k_2^2\mathrm{d}k_2\mathrm{d}\cos(\theta_2)\mathrm{d}\phi_2
=\\ 
&2\pi k_1k_2k_3\mathrm{d}k_1\mathrm{d}k_2\mathrm{d}k_3\mathrm{d}\cos{\theta_1}\mathrm{d}\phi_1\mathrm{d}\xi,
\end{align}
and the factor of $2\pi$ is to ensure that the expectation value of the
estimator matches the definition in Eq.~ (\ref{eq:BispAng}).

The variance of the bispectrum multipoles is then
\begin{align}
    \nonumber   
    &\langle \Delta \overline{B}_{\ell m}(k_1,k_2,k_3)\Delta\overline{B}_{\ell'
m'}(k_1,k_2,k_3) \rangle =\\ \nonumber
&\quad\quad\frac{V_\mathrm{s}}{2\pi}\displaystyle\int \!\mathrm{d}\cos(\theta)\,\mathrm{d}\xi Y^{m
*}_\ell(\theta,\xi)Y^{m' *}_{\ell'}(\theta,\xi) \\
&\hspace{10pt}\times\hspace{-35pt}\quad\quad\quad\quad\left[P(\textbf{k}_1) + \frac{1}{\overline{n}}\right]  \left[P(\textbf{k}_2) +
\frac{1}{\overline{n}}\right]\left[P(\textbf{k}_3) +
\frac{1}{\overline{n}}\right] \label{eq:BispCov}
\end{align}
The derivation of this result is analogous to the power spectrum multipole covariance described in
\citet{yamamoto2006measurement}.

Since we work in the limit of infinitely small $k$-bins only the multipoles with
all $\bm{k}_i$ identical are correlated, but in general there is a correlation
between multipoles with different values of $\ell$ and $m$.

\section{Constraining Cosmological Parameters}
\label{sec:fisher}

For brevity we will use the following notation:

\begin{align}
\mathrm{VarP}_{\bm{k}} &\equiv \langle\left[\Delta P(\bm{k})\right]^2\rangle\\
\mathrm{VarB}_{\bm{k}_1\bm{k}_2} &\equiv \frac{\langle\left[\Delta B(\bm{k}_1\bm{k}_2)\right]^2\rangle}{V_\mathrm{s}}\\
\mathrm{VarB}^{\ell m\ell' m'}_{k_1k_2k_3} &\equiv
\frac{2\pi}{V_\mathrm{s}}\langle\Delta B_{\ell m}(k_1,k_2,k_3)\Delta B_{\ell' m'}(k_1,k_2,k_3)\rangle
\end{align}

\subsection{Information Content of the Full Bispectrum}
We use a Fisher information formalism \citep{tegmark1997measuring, albrecht2006report} to derive expected constraints on
cosmological parameters $\bm{\theta} \equiv (b_1, b_2, f, \alpha_\bot,
\alpha_\parallel)$. 

For the power spectrum we follow the well established procedure of computing
\begin{equation}
\label{eq:Pkfisher}
F_{ij}=\frac{V_\mathrm{s}}{(2\pi)^3}\displaystyle\int\!\mathrm{d}\bm{k}\frac{\partial
P(\bm{k})}{\partial\theta_i}\left(\mathrm{VarP}_{\bm{k}}\right)^{-1}\frac{\partial
P(\bm{k})}{\partial\theta_i}.
\end{equation}
Since the Fourier transform is computed over a finite volume the
$\delta(\bm{k})$ measurements are independent only at discrete points in
$\bm{k}$ space. The density of these points is $V_\mathrm{s}/(2\pi)^3$. The
factor in front of Eq.~\eqref{eq:Pkfisher} renormalizes the continuous integral
over all $\bm{k}$ which would otherwise overestimate the available information.

We numerically compute the integral
\begin{equation}\label{eq:FishSimp}
F_{ij} =
\frac{V_\mathrm{s}}{(2\pi)^2}\displaystyle\int\!k^2\mathrm{d}\cos(\theta)\frac{\partial
P(\bm{k})}{\partial\theta_i}\left(\mathrm{VarP}_{\bm{k}}\right)^{-1}\frac{\partial
P(\bm{k})}{\partial\theta_i},
\end{equation}
where the power spectrum derivatives are obtained by numerically differentiating
Eq.~\eqref{eq:Plin} and the power spectrum variance is given by Eq.~\eqref{eq:Pcov}.
The integration limits are $0 < k < 0.2$ $h/\mathrm{Mpc}$ and $0 < \cos(\theta) <
1$. The first restriction reflects the fact that the statistical properties of
the galaxy field are difficult to model at high wavenumbers because of the
effects of nonlinear evolution and baryonic physics and are usually omitted from
the analysis. The second restriction reflects the fact that a Fourier transform
of a real field obeys $\delta(\bm{k}) = \delta^*(-\bm{k})$ symmetry, which
implies that the power spectrum estimates (which are proportional to
$|\delta(\bm{k})|^2$) are not independent above and below the $z$ axis.
Eq.~\eqref{eq:FishSimp} has one less factor of $2\pi$ compared to Eq.~\eqref{eq:Pkfisher}
because we integrate over azimuthal angle $0< \phi< 2\pi$ on which neither the
power spectrum nor its variance depend.

For the full bispectrum we similarly numerically integrate over all
possible triangles (both the shape and the configuration) and propagate the
information to the cosmological parameters. The Fisher matrix of cosmological
parameters in this case is given by
\begin{equation}
F_{ij}=\frac{V_\mathrm{s}^2}{(2\pi)^6}\displaystyle\int\!\mathrm{d}\bm{k}_1\mathrm{d}\bm{k}_2 
\frac{\partial B(\bm{k}_1\bm{k}_2)}{\partial \theta_i}
\left(V_\mathrm{s}\mathrm{VarB}_{\bm{k}_1,\bm{k}_2}\right)^{-1} \frac{\partial B(\bm{k}_1,\bm{k}_2)}{\partial \theta_j},
\end{equation}
where the factor of $V_\mathrm{s}^2/(2\pi)^6$ accounts for the density of
points on a $k$-grid due to finite volume of the survey, as before.
The integral can be reduced to five dimensions
\begin{align}
\nonumber
F_{ij}=\frac{V_\mathrm{s}}{(2\pi)^5}&\displaystyle\int\!\mathrm{d}k_1\mathrm{d}k_2\mathrm{d}k_3\mathrm{d}\cos(\theta_1)\mathrm{d}\xi\\
&\hspace{-31pt}\times\frac{\partial B(\bm{k}_1\bm{k}_2)}{\partial \theta_i}
\left(\mathrm{VarB}_{\bm{k}_1,\bm{k}_2}\right)^{-1} \frac{\partial B(\bm{k}_1,\bm{k}_2)}{\partial \theta_j},
\end{align}
as the integration over $\phi_1$ azimuthal angle is simply $2\pi$.

We use Eq.~(\ref{eq:bisp}) to compute the bispectrum (and its derivatives) and
Eq.~(\ref{eq:Bcov}) to compute the covariance matrix of the bispectrum.  A
permutation of vectors $\bm{k}_i$ corresponds to the same bispectrum
measurement. In order to account for this symmetry and not double count the data
we impose a condition $k_1>k_2>k_3$ on the integration volume in addition to
$k_i < 0.2\ h/\mathrm{Mpc}$ restriction on each wavevector. We also impose the
triangularity condition $k_1-k_2<k_3$.

\subsection{Information Content of the Multipoles}

The Fisher matrix of cosmological parameters from bispectrum multipoles is given
a three dimensional integral over a sum
\begin{align}
\label{bispinfo}
&F_{ij}=\frac{V_\mathrm{s}^2}{(2\pi)^6}\displaystyle\int\!\mathrm{d}k_1\mathrm{d}k_2\mathrm{d}k_3 \, k_1 k_2 k_3
\\
\nonumber
&\times\displaystyle\sum\limits_{\ell\ell'
mm'}\!\frac{\partial B_{\ell m}(k_1,k_2,k_3) }{\partial \theta_i}
\left(\frac{V_\mathrm{s}}{2\pi}\mathrm{VarB}_{\ell\ell' mm'}^{k_1k_2k_3}\right)^{-1} \frac{\partial B_{\ell' m'}(k_1,k_2,k_3)}{\partial \theta_j},
\end{align}
where the integration is over all possible triangle shapes. Similarly to the 
bispectrum we impose a restriction that $k_1>k_2>k_3$ and that the three sides
satisfy the triangularity condition $k_1-k_2<k_3$. We also restrict ourselves
to triangles with $k_1<0.2 h/\mathrm{Mpc}$.

We use Eq.~ (\ref{eq:BispEst}) to compute numerical derivatives of the multipoles and
Eq.~ (\ref{eq:BispCov}) to compute the variance of the multipoles (and covariance between
them).  We evaluate the sum for increasing values of $\ell_\mathrm{max}$. To
check the effects of higher order terms in $m$ we either take all values of
$-\ell \leq m \leq \ell$ or only the $m=0$. We also try only $m=0$ modes for increasing
even values of $\ell_\mathrm{max}$.

%
%

\section{Results}
\label{sec:results}    
Results in this section are derived assuming a spatially flat $\Lambda CDM$
cosmological model with $\Omega_\mathrm{m}=0.28$, $\Omega_\Lambda=0.72$. 
We consider LRG and ELG samples expected from DESI. For the number density profile
and the bias as a function of redshift we use the same numbers as \cite{tellarini}.

\begin{figure*}
    
    \begin{subfigure}[b]{.4\textwidth}
        \includegraphics[width=\textwidth]{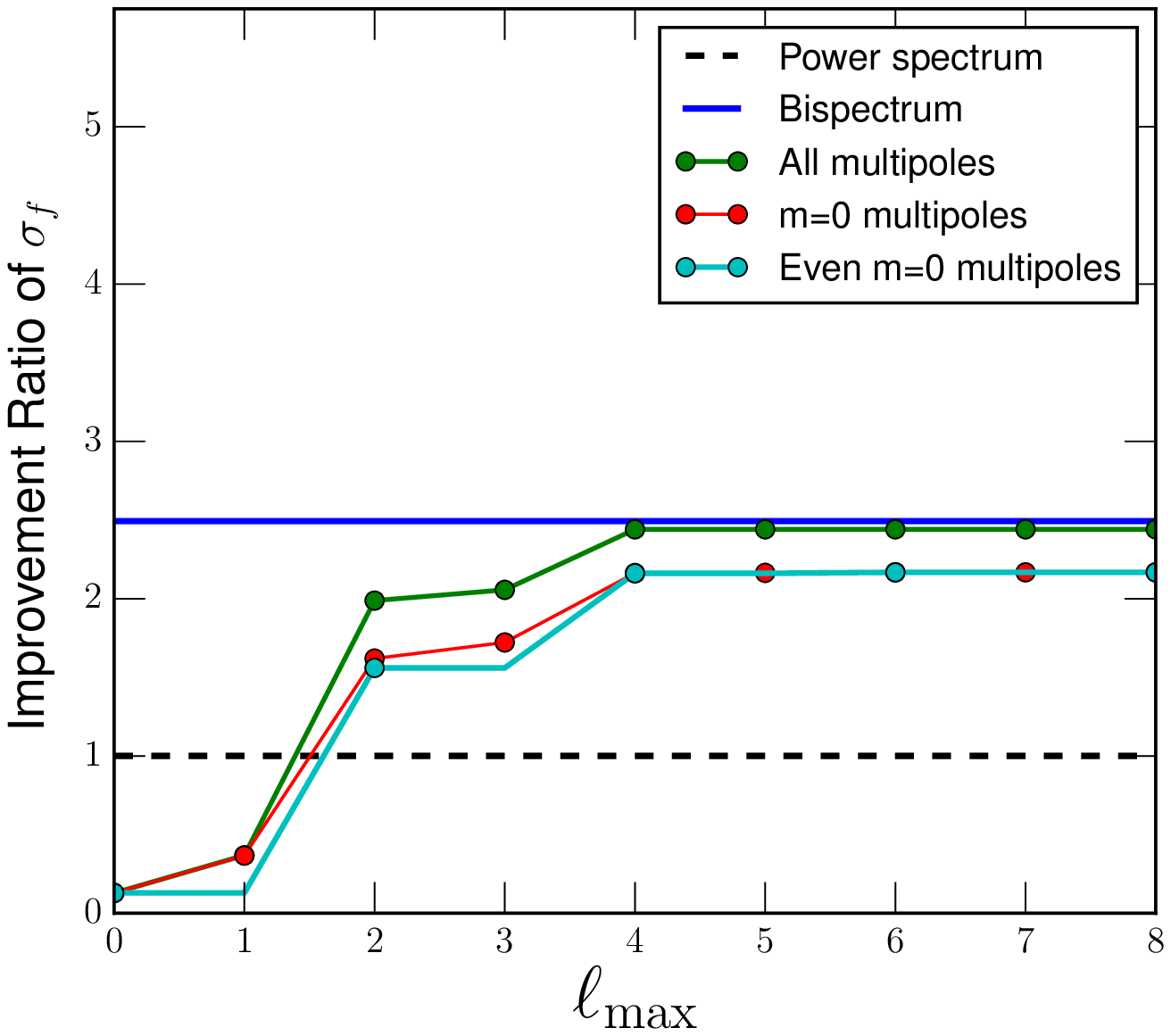}
        \label{fig:f}
    \end{subfigure}
    \begin{subfigure}[b]{0.4\textwidth}
        \includegraphics[width=\textwidth]{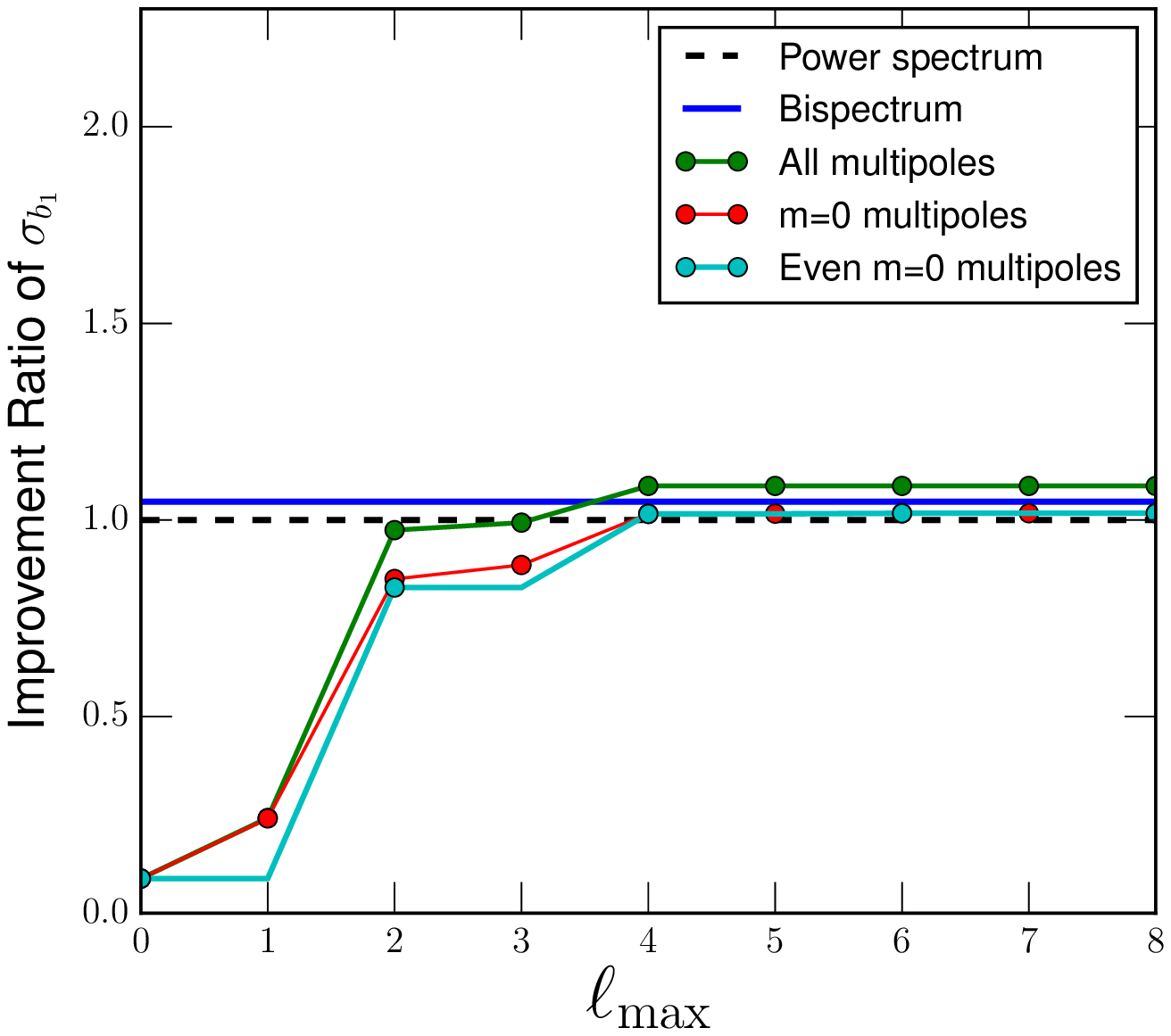}
        \label{fig:b1}
    \end{subfigure}
    ~ 
    
        \begin{subfigure}[b]{0.4\textwidth}
        \includegraphics[width=\textwidth]{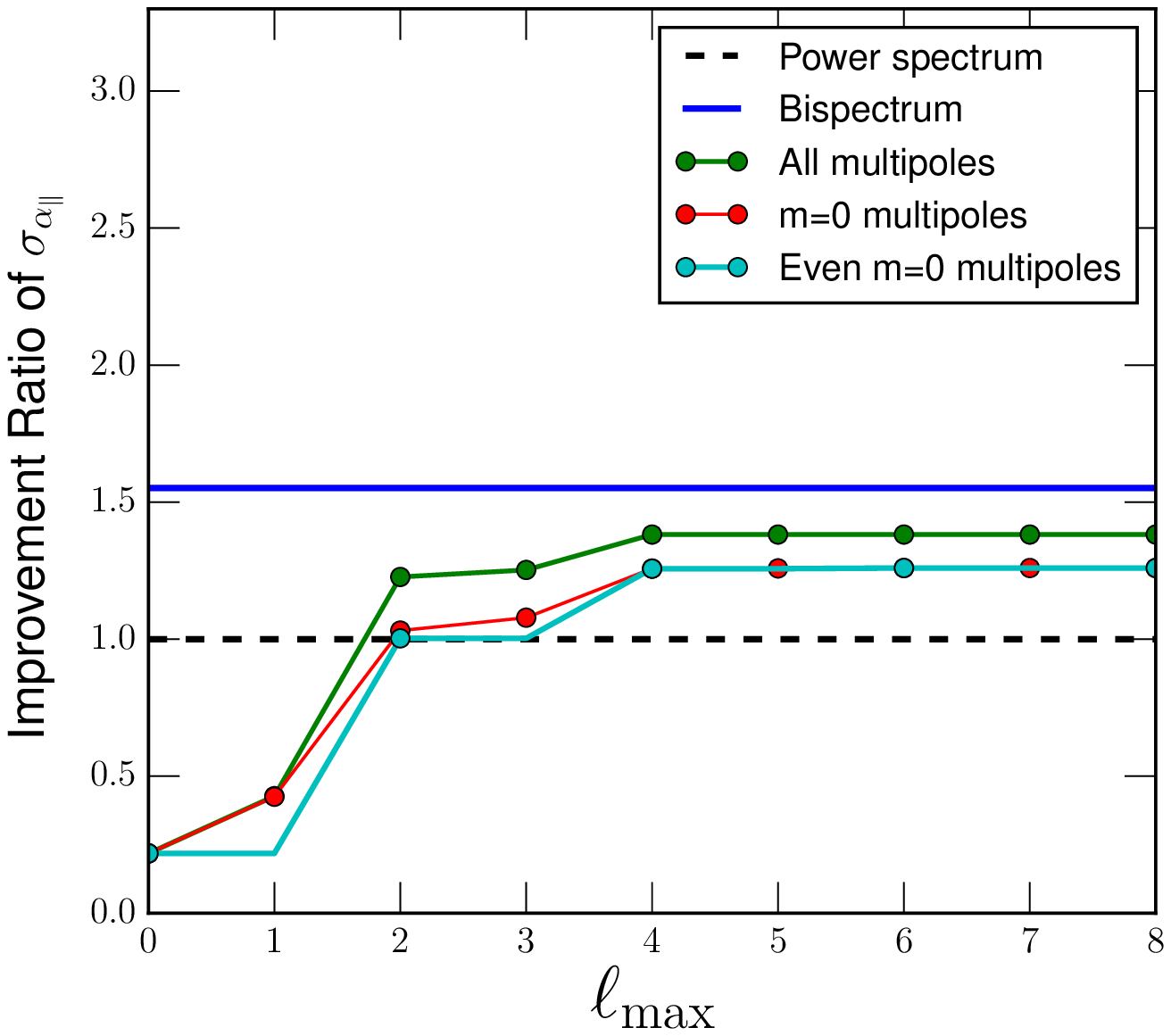}
        \label{fig:apar}
    \end{subfigure}
    ~ 
    \begin{subfigure}[b]{0.4\textwidth}
        \includegraphics[width=\textwidth]{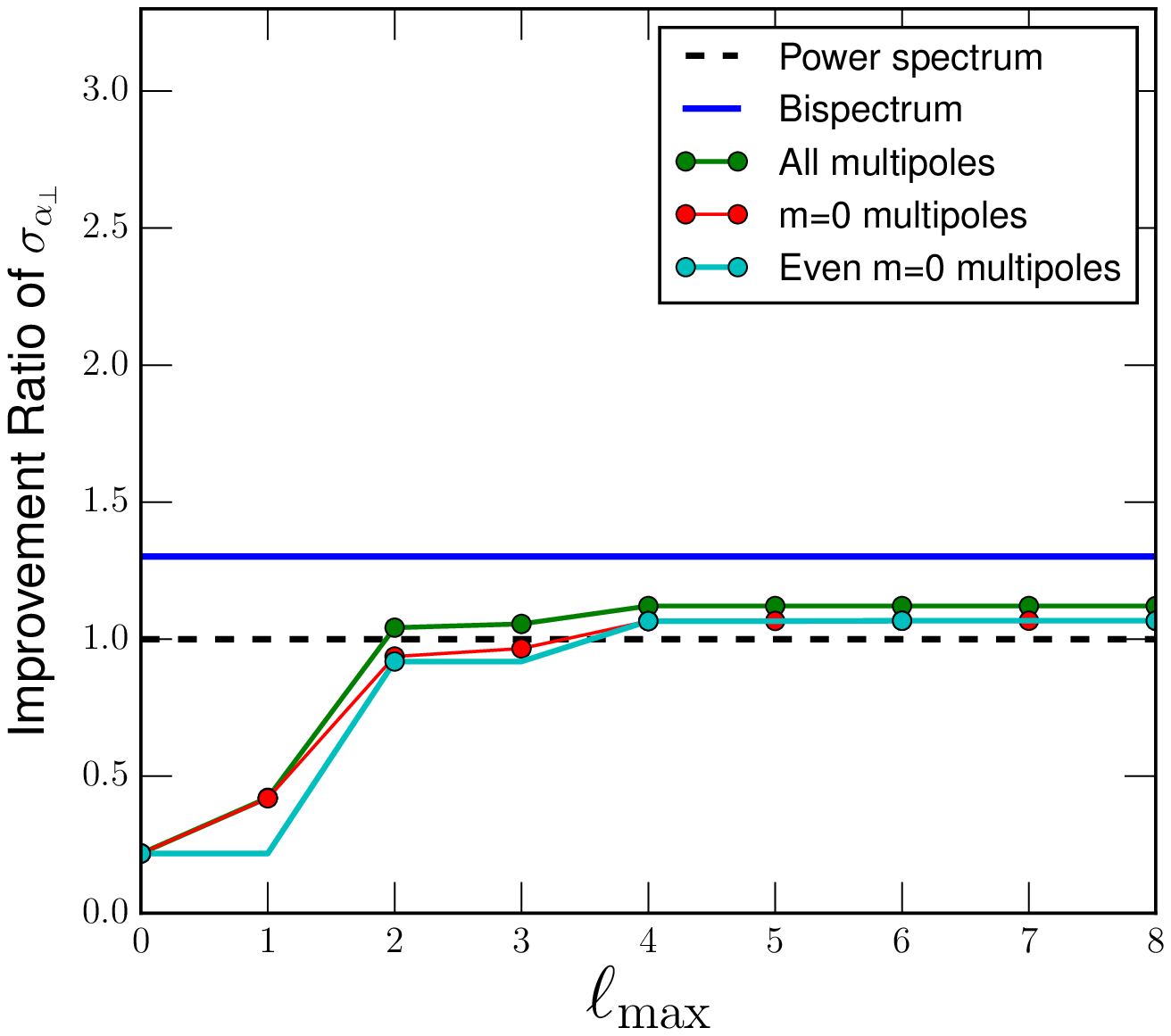}
        \label{fig:aper}
    \end{subfigure}
    \caption{Cosmological constraints expected from the bispectrum multipoles as
a function of maximum $\ell$ used in the analysis for a sample of DESI LRGs in $0.6 < z < 0.7$. The constraints from power
spectrum and the full bispectrum are also displayed for
comparison. The results are normalized to the expected power spectrum
constraints so that the ordinate axis is an improvement factor over the power 
spectrum. The multipole constraints can never be stronger than the full bispectrum constraints. Our top right panel is consistent with this within the numerical error associated with monte carlo integration.}\label{fig:saturation}
\end{figure*}

Fig.~\ref{fig:saturation} shows the expected cosmological constraints on
$\bm{\theta}$ from the bispectrum multipoles for increasing values of
$\ell_\mathrm{max}$. These results are for the LRG sample in the redshift range $0.6 < z
< 0.7$, We compute this for all $\ell$ and $m$ values, all $\ell$ values with
only $m=0$, and for only even $\ell$ modes with $m=0$. We show expected
constraints from the power spectrum and the bispectrum on the same plots for
comparison. 

Fig.~\ref{fig:saturation} shows that the full (unreduced) bispectrum is capable of providing better constraints compared to the power spectrum. This is
especially true for the growth rate parameter $f$ where the improvement is
almost a factor of 2 in the statistical errors. For the  $\alpha$ parameters the constraints derived 
from the full bispectrum are still a factor of about 1.5 better compared to the
power spectrum, but become slightly worse for the multipoles. In all cases the information in the
multipoles seems to be mostly in the first three even $\ell$ modes with $m=0$.

\begin{figure*}

    \begin{subfigure}[b]{0.4\textwidth}
        \includegraphics[width=\textwidth]{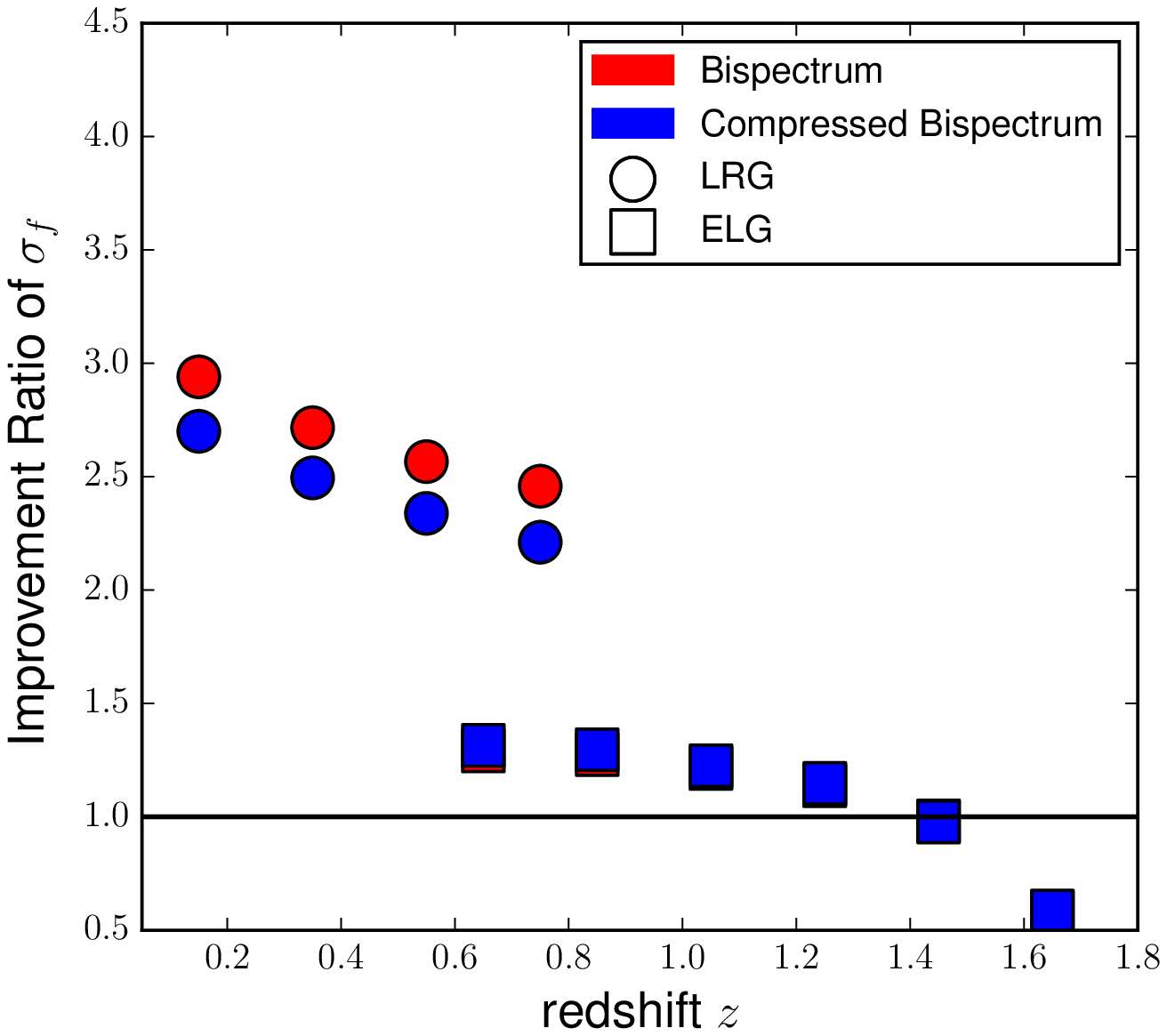}
        \label{fig:fz}
    \end{subfigure}
    ~ 
    \begin{subfigure}[b]{0.4\textwidth}
        \includegraphics[width=\textwidth]{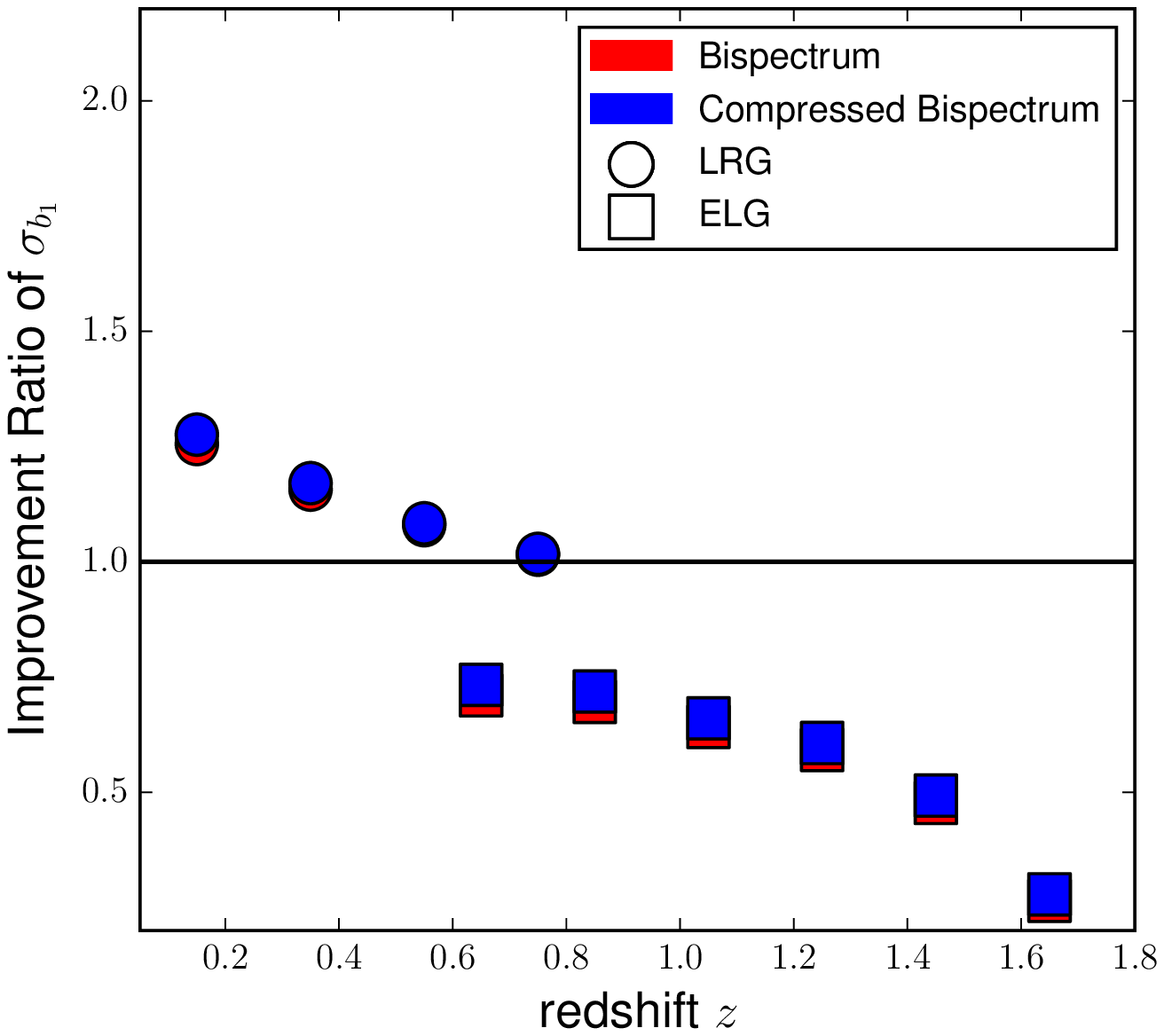}
        \label{fig:b1z}
    \end{subfigure}
    
        \begin{subfigure}[b]{0.4\textwidth}
        \includegraphics[width=\textwidth]{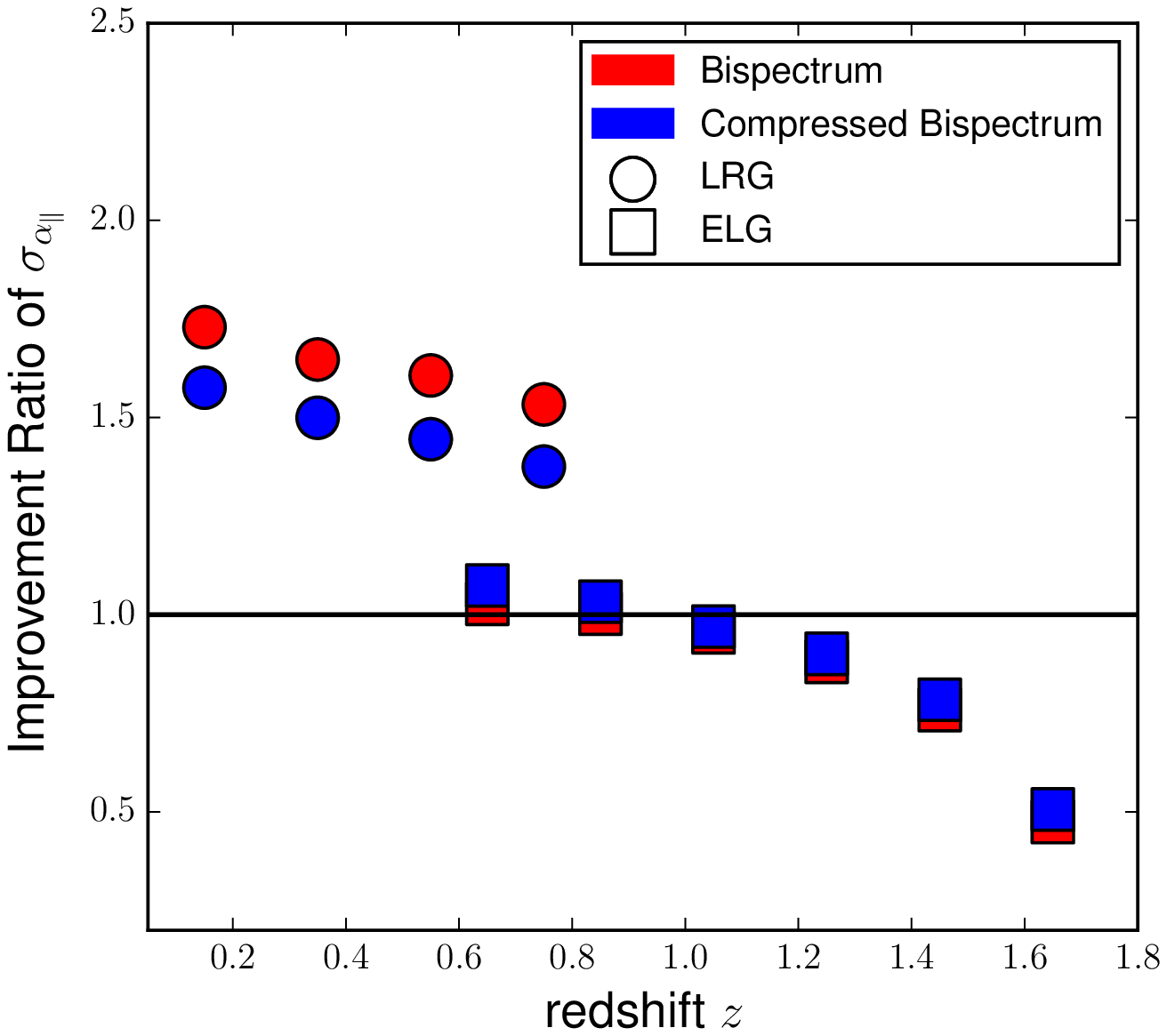}
        \label{fig:aparz}
    \end{subfigure}
    ~ 
    \begin{subfigure}[b]{0.4\textwidth}
        \includegraphics[width=\textwidth]{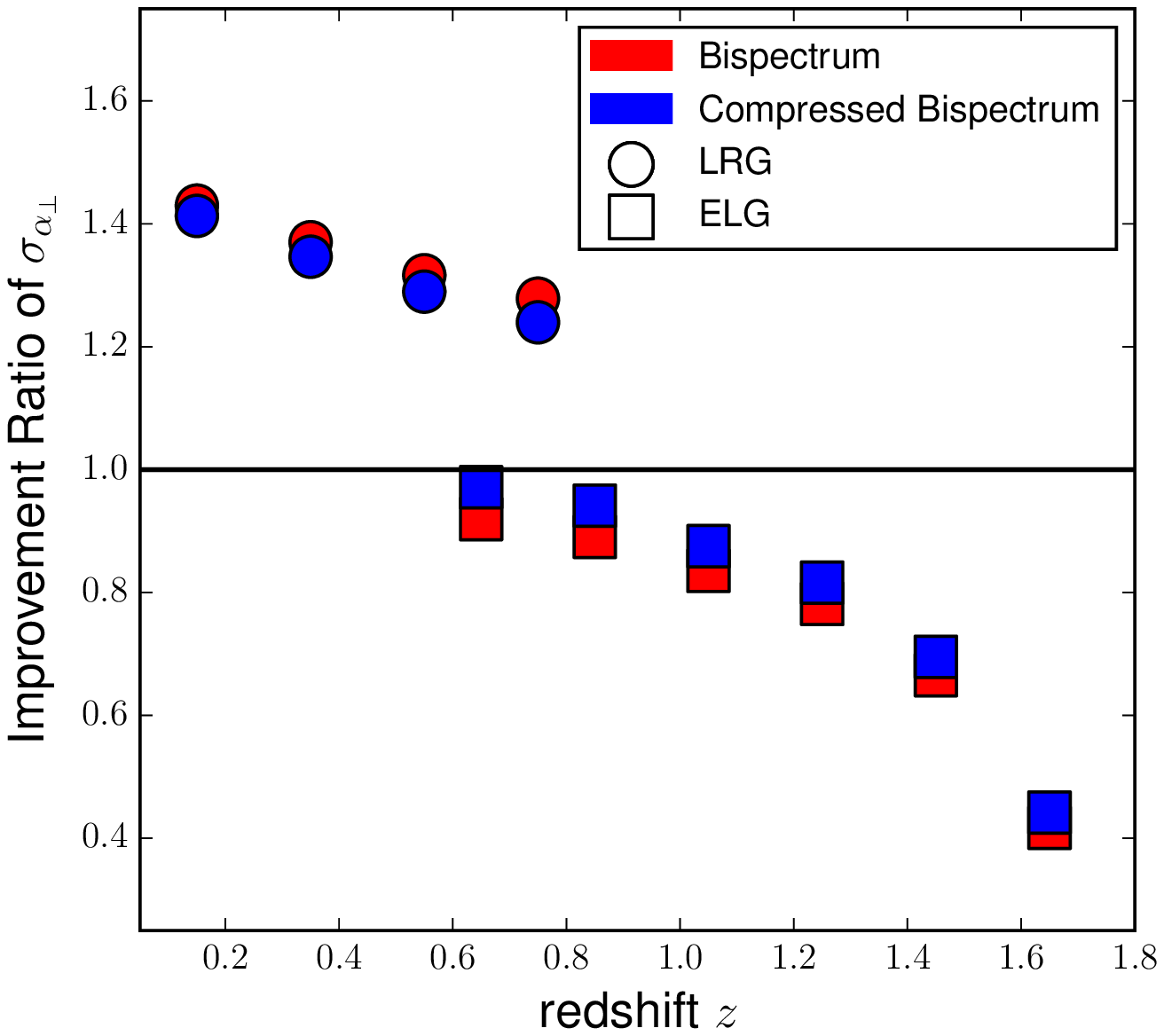}
        \label{fig:aperz}
    \end{subfigure}
    \caption{Improvement on derived errors of cosmological
parameters compared to the power spectrum for different redshifts and tracer types. Red symbols (on top)
represent the constraints derivable from the full bispectrum, while the blue
symbols (on the bottom) represent constraints from first three even multipoles
with $m=0$. For some redshifts the multipole constraints are slightly better than the full bispectrum constraints, but they are consistent within the numerical errors associated with the monte carlo integration.}\label{fig:improvement}
\end{figure*}

The behaviour seems to be qualitatively similar for other redshifts and tracers.
Fig.~\ref{fig:improvement} shows similar results over a wider redshift range.
This means that the first even multipoles averaged over azimuthal angle are as
good as the full bispectrum for the purposes of deriving cosmological constraints.

The bispectrum provides significantly larger improvement over power spectrum at
low redshifts. This is due to a high number density of galaxies and the higher 
amplitude of fluctuations.

%
%

\section{Conclusions}
\label{sec:conclusions}

We developed a Fisher information matrix based method of computing the expected
constraints on cosmological parameters from the bispectrum and the angular
multipoles of the bispectrum of a given galaxy sample. Since the full bispectrum
is difficult to analyse, some kind of data reduction will inevitable have to be
applied to the measurements. We computed the information loss associated with
the commonly proposed reduction schemes that rely on angular integration of
the bispectrum.

We find that the full bispectrum alone can deliver cosmological constraints that 
are a factor of few better than the ones derivable from the power spectrum at low $z$. The
improvement is especially large for the growth rate parameter $f$ where the
improvement on the measurement error is almost a factor of 3.
The improvement is the largest at lower redshifts where the number density of
galaxies in the sample is the highest. Most of the information is in the first
three even multipoles with $m=0$, which means that just three numbers per
bispectrum shape are enough for the purposes of obtaining cosmological
constraints.

Our results at first may seem to contradict previously published results
that claim a more modest improvement when adding the bispectrum to the power spectrum
\citep{sefusattikomatsu,Szapudibook,CarronNeyrinck,CarronSzapudi}.
This is due to a number of reasons. Many previous works have looked at the monopole
of the bispectrum which will obviously contain much less information on $f$. 
The bispectrum information increases more steeply compared to the power spectrum
with the number density of galaxies, therefore this large improvement will only
result in future dense surveys and will not necessarily show in current and past
surveys that have a lower galaxy number density. Finally, many past claims refer to
``amplitude like'' parameters (e.g. primordial amplitude of fluctuations) for
isotropic fields. The $f$ parameter is not really ``amplitude like'' since it describes
an angular dependent variations in the statistics, and the 5D shape of the bispectrum
turns out to be more sensitive to this parameter than it would be to a mere change in
amplitude.

Our results are consistent with the ones reported in \cite{song2015cosmology} if
we only consider strictly linear scales of $k_i < 0.1\ \mathrm{Mpc/h}$. This is
expected since the bispectrum signal to noise scales better with increasing $k_\mathrm{max}$ compared to the power spectrum.
Their model includes the Finger of God effects and therefore the forecasts are more conservative and realistic.
Since our main goal was not to produce accurate forecasts but rather to study the effects of the multipole reduction  we decided to sacrifice the realism of constraints for clarity. We explicitly checked that our main conclusions are robust with respect to the choice of $k_\mathrm{max}$ and do not change when we include $\sigma_\mathrm{FOG}$.

In this work we do not consider a cross correlation between the power spectrum
and the bispectrum measurements and it is difficult to say how big the overall
improvement in the errors is when the two are properly combined \citep[see][for correlated full bispectrum DESI forecasts]{song2015cosmology}. We know however
that the improvement will be at least as big as the improvement from the
bispectrum (or the bispectrum multipoles) alone. Recent studies indicated that
the cosmological constraints from power spectrum and bispectrum are not very
strongly correlated \citep{slepian2016modeling, slepian2016detection, gil2016clustering}, so the improvement may actually be much larger.

The main conclusions from our work are as follows:
\begin{itemize}
\item{The bispectrum measurements from future surveys have a potential of
improving the growth rate measurements by at least a factor of 2.5 at low redshifts (this is a very conservative estimate assuming that the bispectrum information is perfectly correlated with the power spectrum).}
\item{When expanding the bispectrum in angular multipoles, the three numbers
corresponding to the first three even terms with $m=0$ in the multipole
expansion contain most of the information relevant for the derivation of
cosmological constraints.}
\end{itemize}

\section*{Acknowledgements}

We thank H\'{e}ctor Gil Marin, Florian Beutler, Eichiro Komatsu, Cristiano Porciani, Emiliano Sefusatti and David Pearson for useful discussions. 
This work was supported by SNSF grant SCOPES IZ73Z0 152581, GNSF grant FR/339/6-350/14, and 
NASA grant 12-EUCLID11-0004. This work was supported in part by DOE grant DEFG 03-99EP41093.
We have used NASA's Astrophysics Data System Bibliographic Service and the arXiv e-print 
service for bibliography search and \text{http://cosmocalc.icrar.org/} for computating some 
cosmological parameters.

\bibliographystyle{mnras}

\bibliography{bibli.bib}

\begin{thebibliography}{}
\makeatletter
\relax
\def\mn@urlcharsother{\let\do\@makeother \do\$\do\&\do\#\do\^\do\_\do\%\do\~}
\def\mn@doi{\begingroup\mn@urlcharsother \@ifnextchar [ {\mn@doi@}
  {\mn@doi@[]}}
\def\mn@doi@[#1]#2{\def\@tempa{#1}\ifx\@tempa\@empty \href
  {http://dx.doi.org/#2} {doi:#2}\else \href {http://dx.doi.org/#2} {#1}\fi
  \endgroup}
\def\mn@eprint#1#2{\mn@eprint@#1:#2::\@nil}
\def\mn@eprint@arXiv#1{\href {http://arxiv.org/abs/#1} {{\tt arXiv:#1}}}
\def\mn@eprint@dblp#1{\href {http://dblp.uni-trier.de/rec/bibtex/#1.xml}
  {dblp:#1}}
\def\mn@eprint@#1:#2:#3:#4\@nil{\def\@tempa {#1}\def\@tempb {#2}\def\@tempc
  {#3}\ifx \@tempc \@empty \let \@tempc \@tempb \let \@tempb \@tempa \fi \ifx
  \@tempb \@empty \def\@tempb {arXiv}\fi \@ifundefined
  {mn@eprint@\@tempb}{\@tempb:\@tempc}{\expandafter \expandafter \csname
  mn@eprint@\@tempb\endcsname \expandafter{\@tempc}}}

\bibitem[\protect\citeauthoryear{Ade et~al.,}{Ade et~al.}{2014}]{ade2014planck}
Ade P.,  et~al., 2014, Astronomy \& Astrophysics, 571, A16

\bibitem[\protect\citeauthoryear{Albrecht et~al.,}{Albrecht
  et~al.}{2006}]{albrecht2006report}
Albrecht A.,  et~al., 2006, arXiv preprint astro-ph/0609591

\bibitem[\protect\citeauthoryear{Alcock \& Paczynski}{Alcock \&
  Paczynski}{1979}]{alcock1979evolution}
Alcock C.,  Paczynski B.,  1979, Nature, 281, 358

\bibitem[\protect\citeauthoryear{Ballinger, Peacock  \& Heavens}{Ballinger
  et~al.}{1996}]{ballinger1996measuring}
Ballinger W.,  Peacock J.,   Heavens A.,  1996, arXiv preprint astro-ph/9605017

\bibitem[\protect\citeauthoryear{Beutler et~al.,}{Beutler
  et~al.}{2014}]{beutler2014clustering}
Beutler F.,  et~al., 2014, Monthly Notices of the Royal Astronomical Society,
  443, 1065

\bibitem[\protect\citeauthoryear{{Carron} \& {Neyrinck}}{{Carron} \&
  {Neyrinck}}{2012}]{CarronNeyrinck}
{Carron} J.,  {Neyrinck} M.~C.,  2012, \mn@doi [\apj]
  {10.1088/0004-637X/750/1/28}, \href
  {http://adsabs.harvard.edu/abs/2012ApJ...750...28C} {750, 28}

\bibitem[\protect\citeauthoryear{{Carron} \& {Szapudi}}{{Carron} \&
  {Szapudi}}{2014}]{CarronSzapudi}
{Carron} J.,  {Szapudi} I.,  2014, \mn@doi [\mnras] {10.1093/mnrasl/slt167},
  \href {http://adsabs.harvard.edu/abs/2014MNRAS.439L..11C} {439, L11}

\bibitem[\protect\citeauthoryear{Feldman, Kaiser  \& Peacock}{Feldman
  et~al.}{1993}]{feldman1993power}
Feldman H.~A.,  Kaiser N.,   Peacock J.~A.,  1993, arXiv preprint
  astro-ph/9304022

\bibitem[\protect\citeauthoryear{Gil-Mar{\'\i}n, Nore{\~n}a, Verde, Percival,
  Wagner, Manera  \& Schneider}{Gil-Mar{\'\i}n et~al.}{2015}]{gil2015power}
Gil-Mar{\'\i}n H.,  Nore{\~n}a J.,  Verde L.,  Percival W.~J.,  Wagner C.,
  Manera M.,   Schneider D.~P.,  2015, Monthly Notices of the Royal
  Astronomical Society, 451, 5058

\bibitem[\protect\citeauthoryear{Gil-Mar{\'\i}n, Percival, Verde, Brownstein,
  Chuang, Kitaura, Rodr{\'\i}guez-Torres  \& Olmstead}{Gil-Mar{\'\i}n
  et~al.}{2016}]{gil2016clustering}
Gil-Mar{\'\i}n H.,  Percival W.~J.,  Verde L.,  Brownstein J.~R.,  Chuang
  C.-H.,  Kitaura F.-S.,  Rodr{\'\i}guez-Torres S.~A.,   Olmstead M.~D.,  2016,
  arXiv preprint arXiv:1606.00439

\bibitem[\protect\citeauthoryear{Greig, Komatsu  \& Wyithe}{Greig
  et~al.}{2013}]{greig2013cosmology}
Greig B.,  Komatsu E.,   Wyithe J. S.~B.,  2013, Monthly Notices of the Royal
  Astronomical Society, 431, 1777

\bibitem[\protect\citeauthoryear{Jackson}{Jackson}{1972}]{jackson1972critique}
Jackson J.,  1972, Monthly Notices of the Royal Astronomical Society, 156, 1P

\bibitem[\protect\citeauthoryear{Kaiser}{Kaiser}{1987}]{kaiser1987clustering}
Kaiser N.,  1987, Monthly Notices of the Royal Astronomical Society, 227, 1

\bibitem[\protect\citeauthoryear{Kazin et~al.,}{Kazin
  et~al.}{2010}]{kazin2010baryonic}
Kazin E.~A.,  et~al., 2010, The Astrophysical Journal, 710, 1444

\bibitem[\protect\citeauthoryear{Laureijs et~al.,}{Laureijs
  et~al.}{2011}]{laureijs2011euclid}
Laureijs R.,  et~al., 2011, arXiv preprint arXiv:1110.3193

\bibitem[\protect\citeauthoryear{Levi et~al.,}{Levi
  et~al.}{2013}]{levi2013desi}
Levi M.,  et~al., 2013, arXiv preprint arXiv:1308.0847

\bibitem[\protect\citeauthoryear{Peebles}{Peebles}{1980}]{peebles1980large}
Peebles P. J.~E.,  1980, The large-scale structure of the universe.
Princeton university press

\bibitem[\protect\citeauthoryear{Samushia et~al.,}{Samushia
  et~al.}{2011}]{samushia2011effects}
Samushia L.,  et~al., 2011, Monthly Notices of the Royal Astronomical Society,
  410, 1993

\bibitem[\protect\citeauthoryear{Schlegel et~al.,}{Schlegel
  et~al.}{2009}]{schlegel2009bigboss}
Schlegel D.~J.,  et~al., 2009, arXiv preprint arXiv:0904.0468

\bibitem[\protect\citeauthoryear{Scoccimarro}{Scoccimarro}{2000}]{scoccimarro2000bispectrum}
Scoccimarro R.,  2000, The Astrophysical Journal, 544, 597

\bibitem[\protect\citeauthoryear{Scoccimarro}{Scoccimarro}{2015}]{scoccimarro2015fast}
Scoccimarro R.,  2015, Physical Review D, 92, 083532

\bibitem[\protect\citeauthoryear{Scoccimarro, Couchman  \& Frieman}{Scoccimarro
  et~al.}{1999}]{scoccimarro1999bispectrum}
Scoccimarro R.,  Couchman H.,   Frieman J.~A.,  1999, The Astrophysical
  Journal, 517, 531

\bibitem[\protect\citeauthoryear{{Sefusatti} \& {Komatsu}}{{Sefusatti} \&
  {Komatsu}}{2007}]{sefusattikomatsu}
{Sefusatti} E.,  {Komatsu} E.,  2007, \mn@doi [\prd]
  {10.1103/PhysRevD.76.083004}, \href
  {http://adsabs.harvard.edu/abs/2007PhRvD..76h3004S} {76, 083004}

\bibitem[\protect\citeauthoryear{Sefusatti, Crocce, Pueblas  \&
  Scoccimarro}{Sefusatti et~al.}{2006}]{sefusatti2006cosmology}
Sefusatti E.,  Crocce M.,  Pueblas S.,   Scoccimarro R.,  2006, Physical Review
  D, 74, 023522

\bibitem[\protect\citeauthoryear{Sefusatti, Crocce  \& Desjacques}{Sefusatti
  et~al.}{2012}]{sefusatti2012halo}
Sefusatti E.,  Crocce M.,   Desjacques V.,  2012, Monthly Notices of the Royal
  Astronomical Society, 425, 2903

\bibitem[\protect\citeauthoryear{Simpson \& Peacock}{Simpson \&
  Peacock}{2010}]{simpson2010difficulties}
Simpson F.,  Peacock J.~A.,  2010, Physical Review D, 81, 043512

\bibitem[\protect\citeauthoryear{Slepian \& Eisenstein}{Slepian \&
  Eisenstein}{2016}]{slepian2016modeling}
Slepian Z.,  Eisenstein D.~J.,  2016, arXiv preprint arXiv:1607.03109

\bibitem[\protect\citeauthoryear{Slepian et~al.,}{Slepian
  et~al.}{2016}]{slepian2016detection}
Slepian Z.,  et~al., 2016, arXiv preprint arXiv:1607.06097

\bibitem[\protect\citeauthoryear{Song, Taruya  \& Oka}{Song
  et~al.}{2015}]{song2015cosmology}
Song Y.-S.,  Taruya A.,   Oka A.,  2015, Journal of Cosmology and Astroparticle
  Physics, 2015, 007

\bibitem[\protect\citeauthoryear{{Szapudi}}{{Szapudi}}{2009}]{Szapudibook}
{Szapudi} I.,  2009, in {Mart{\'{\i}}nez} V.~J.,  {Saar} E.,
  {Mart{\'{\i}}nez-Gonz{\'a}lez} E.,   {Pons-Border{\'{\i}}a} M.-J.,  eds,
  Lecture Notes in Physics, Berlin Springer Verlag Vol. 665, Data Analysis in
  Cosmology. pp 457--492, \mn@doi{10.1007/978-3-540-44767-2_14}

\bibitem[\protect\citeauthoryear{Taruya, Saito  \& Nishimichi}{Taruya
  et~al.}{2011}]{taruya2011forecasting}
Taruya A.,  Saito S.,   Nishimichi T.,  2011, Physical Review D, 83, 103527

\bibitem[\protect\citeauthoryear{Taylor \& Hamilton}{Taylor \&
  Hamilton}{1996}]{taylor1996non}
Taylor A.,  Hamilton A.,  1996, Monthly Notices of the Royal Astronomical
  Society, 282, 767

\bibitem[\protect\citeauthoryear{Tegmark}{Tegmark}{1997}]{tegmark1997measuring}
Tegmark M.,  1997, Physical Review Letters, 79, 3806

\bibitem[\protect\citeauthoryear{{Tellarini}, {Ross}, {Tasinato}  \&
  {Wands}}{{Tellarini} et~al.}{2016}]{tellarini}
{Tellarini} M.,  {Ross} A.~J.,  {Tasinato} G.,   {Wands} D.,  2016, \mn@doi
  [\jcap] {10.1088/1475-7516/2016/06/014}, \href
  {http://adsabs.harvard.edu/abs/2016JCAP...06..014T} {6, 014}

\bibitem[\protect\citeauthoryear{Yamamoto, Nakamichi, Kamino, Bassett  \&
  Nishioka}{Yamamoto et~al.}{2006}]{yamamoto2006measurement}
Yamamoto K.,  Nakamichi M.,  Kamino A.,  Bassett B.~A.,   Nishioka H.,  2006,
  Publications of the Astronomical Society of Japan, 58, 93

\makeatother
\end{thebibliography}

\label{lastpage}

 \end{document}